\documentclass[10 pt]{article}

\usepackage{graphicx,amsmath,amssymb}
\usepackage{hyperref}
\usepackage{graphicx}
\usepackage{xcolor}

\title{ Virtual Black Holes from Generalized Uncertainty Principle and Proton Decay}

\author{Salwa Alsaleh$^{1}$, Abeer Al-Modlej$^1$ and Ahmed Farag Ali $^{2,3}$ }

\begin{document}

\maketitle
\begin{center}
 $^1$ Department of Physics and Astronomy,\\ King Saud University, Riyadh 11451, Saudi Arabia\\
 $^2$ Netherlands Institute for Advanced Study,\\ Korte Spinhuissteeg 3, 1012 CG Amsterdam, Netherlands\\
 $^3$ Department of Physics, Faculty of Science,\\ Benha University, Benha, 13518, Egypt\\
\end{center}

\begin{abstract}
We investigate the formation of virtual black holes in the context of generalized uncertainty principle (GUP), as a mediator for a proton decay process which is forbidden by the standard model. Then, we calculate the lower bounds of the GUP deformation parameter by the experimental bound on the half life of the proton. 
\end{abstract}

\section{Introduction}
The Plank-scale physics is considered as the main focus of quantum gravity approaches. This is because the merge between the quantum uncertainty that predicts an enormous fluctuations of energy at small scales, and the theory of general relativity that associates energy with spacetime curvature leads to the notion of spacetime foam, at the Plank scale \cite{hawking1978spacetime,garattini2002space}. This foam has been investigated in several models of quantum gravity, like the spin foam models in loop quantum gravity (LQG) \cite{perez2003spin,baez1998spin}, $ S^2 \times S^2$ bubbles of `virtual' black holes as described by Hawking \cite{hawking1996virtual}, or fluctuating geometry as described by group field theory (GFT) and String field theory (SFT) \cite{faizal2012some,faizal2015deformation}. The different models of quantum gravity agree that the structure of spacetime at the small scale is significantly different from the smooth structure described by general relativity \cite{doplicher1995quantum}. However, there is disagreement at the particular scale of which this `fuzziness'  in the spacetime start to be prominent. This is mainly due to the disagreement on whether there exist  compactified, or Large extra dimensions as predicted by string theory and Randall-Sundrum model \cite{randall1999large,randall2005book}, respectively or there is no need for such assumption as in LQG, GFT and Hawking bubble model \cite{nicolai2005loop}. 
The existence of a minimal length scale that commonly predicted by various approaches to quantum gravity is manifested phenomenologically by deformation of the standard momentum dispersion relations \cite{cortes2005quantum}
to incorporate a cut-off length $ \ell_p$ - or equivalently- energy $ E_p$ scales \cite{kowalski2005introduction}. This can be achieved by deformation on the metric, as in gravity's rainbow, by the rainbow functions \cite{magueijo2004gravity}, or deformation of Heisenberg algebra \cite{maggiore1993algebraic}. The latter is known as the Generalized uncertainty principle (GUP) \cite{maggiore1993generalized,maggiore1994quantum},  which is based on deforming the commutation relation between momentum and position operators in quantum mechanics, but keeping the associative structure ( Lie algebra structure) of the original commutation relations. The most general type of deformation is \cite{ali2009discreteness}
\begin{align}
[x^{\mu}, x^{\nu}] & =0 & [p_{\mu}, p_{\nu}] &=0 ,
\end{align}
\begin{equation}
[x^{\mu}, p_{\nu}] =  \frac{\hbar}{2}\eta^\mu _{\;\nu} \left\lbrace  1- \frac{\alpha}{E_p}\left( |p|\eta^\mu _{\;\nu}+\frac{p^\mu p_\nu}{|p|}\right) + \frac{\alpha}{E_p}\left( p^2 \eta^\mu _{\;\nu} +3 p^\mu p_\nu \right) \right\rbrace .
\end{equation}
Which leads to the generalized uncertainty relation  
\begin{equation}
\Delta x^\mu \Delta p_\mu \geq  \frac{\hbar}{2}\left[ 1+ \left( \frac{\alpha}{ E_p\,\sqrt{\langle p^2 \rangle}}+4 \frac{\alpha^2}{E_p^2}\right) \Delta p^2 + 4 \frac{\alpha^2}{E_p^2} \langle p\rangle ^2 -2\frac{ \alpha}{E_p} \sqrt{\langle p^2 \rangle}\right] 
\label{GUP}
\end{equation}
The GUP introduces two terms to the standard uncertainty principle, one is linear and the other is quadratic in momentum. This uncertainty relation is invariant under the Doubly-Special relativity transformation, that preserves both the speed of light and the maximum energy $E_p$ \cite{cortes2005quantum,magueijo2002lorentz,magueijo2005string}. It was shown that GUP implies discreteness of space at short distance scale scale, as predicted by LQG \cite{ali2010generalized,cortes2005quantum,ali2009discreteness}. 

The existence of virtual black holes at the scale of quantum gravity is a possible observation of proton decay \cite{adams2001proton}, a process that is forbidden by the standard model due to conservation of baryon number $B$. The proton half life is estimated to be larger than $ \sim 10^{34}$ years, by experimental observations \cite{PhysRevLett.102.141801}. In Quantum gravity approaches, the proton decay is considered as mediator for the virtual black hole. Simply, consider the proton as a spherical object of radius $ r_{proton} \sim 10^{-15}$ m, and virtual black holes form inside of that space ( that is $ 10^{20}$ larger), two of the three point-like quarks could fall into the black hole, to evaporate away only conserving the energy, charge and angular momentum of the two quarks, due to the no hair theorem \cite{barrow1987cosmic}. The decay products could vary, rarely  they could  the same quarks that fell into the black hole. However, in most cases it could be other types of particles.
 proton half-life for this process is calculating from considering the probability of the two quarks to fall into the black hole before it evaporates given the proton crossing time $ \sim M_{proton}^{-1}\sim 10^{-31}$ years, and considering the proton as a hard sphere of radius $ \sim m_p^{-1}$, hence the probability for two quarks to be confined within an region of Plank volume is given by $ \sim (M_{proton}/M_{qg})^3$. Given the fact that virtual black holes decay in Plank time, gives another factor of $ mp/M_{qg}$. Thus,  the proton half life is therefore~\cite{adams2001proton},
\begin{equation}
 \tau_p \sim M_{proton}^{-1} (\frac{M_{qg}}{M_{proton}})^4
 \label{hl}
\end{equation}
{Where $M_{qg}$ is defined to be the minimal mass of virtual black holes predicted by a particular quantum gravity model.} 
It is observed that the proton half-life depends on the quantum gravity mass of the virtual black hole. In the above analysis, the mass $M_{qg} $ is equal to the Planck mass $ M_p$. Thus, using the equation~\eqref{hl}  the proton half life due to this process in the order of $ \sim 10^{45}$ years.
Proton decay via virtual black holes was also studied in large extra dimensions existed, like in Randal-Strudum model. In this case the half life is changed, because of these extra  dimensions, and the probability will change because there is more ` space' for the quarks to move in.
\begin{equation}
 \tau_p \sim M_{proton}^{-1} (\frac{M_{qg}}{M_{proton}})^D
\end{equation}

	The quantum gravity mass $M_{qg}$ here may not be the Plank mass, but it is bounded to be $ M_{qg} >(M^D_p \Lambda^4)^{1/D}$, where $ \Lambda$ is the energy scale defined by experimental bound $ \Lambda \sim 10^{16} GeV$ \cite{dimopoulos2001black}. \\
However, some theories beyond the standard model, like  grand unified theories (GUTs),  supersymmetry (SUSY), electroweak sphaleron anomaly \cite{arnold1987sphalerons}  and magnetic monopoles break the  baryon number conservation.
 \begin{figure}
 	\centering
 	\includegraphics[width = 0.8 \linewidth]{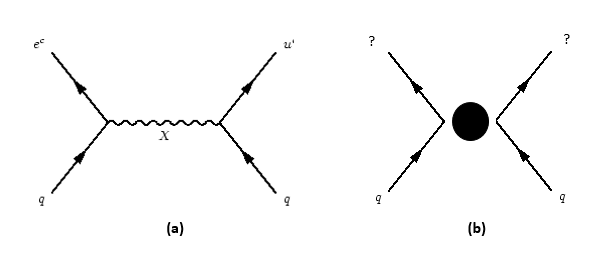}
 	\caption{ Proton decay Feynman diagram (a) Via the X-boson exchange particle, a decay predicted by GUP models. (b) Via two quarks falling into a virtual black hole.  }
 \end{figure}
  Predicting a half life for the proton of $ \sim 10^{30} (\frac{M_X}{10^{15} GeV})^4  $ years for $SU(5) $ Georgi-Glashow  \cite{georgi1974unity}, and $SO(10)$ GUT models \cite{baez2010algebra} models, that predict the decay mediated by a massive X-boson with mass $ M_X > 10^{15} GeV$, making the half life at least $ 10^{32}$ years \cite{langacker1981grand}. SUSY models predict a proton half life close the experimental limit of  $ \sim M^{-2}_{SUSY}$ approximately  $10^{36}-10^{39}$ years \cite{dimopoulos1982proton,bajc2002proton}. Proton decays via  Higgs sector in $ SU(5)$ or $SU(15)$ GUT and magnetic monopoles are also predicted to have a close half life than SUSY (in 4 or higher dimensions) \cite{masiero1982naturally,sakai1982proton,frampton1990higgs}. \begin{table}[h] \label{protonHL}
  	\centering
  	\begin{tabular}{|c|c|}
  		\hline
  		Theory& Proton half life in years $ (\tau_p)$\\
  		\hline
  		Quantum gravity in D=4 & $ \sim 10^{45}$ \\
  		Quantum gravity in $D>4$ & $ \sim 10^{33} 10^{64} (\frac{M_{qg}}{\Lambda})^4$ \\
  		Georgi-Glashow SU(5) & $ \sim   10^{30}-10^{31}$ \\
  		Mimimal SUSY  SU(5)& $ \sim 10^{28}-10^{32} $\\
  		SUSY (MSSM) SU(5)& $ \sim 10^{34} $\\
  		SUSY (D=5) SU(5)& $ \sim 10^{35} $\\
  		SO(10) GUT & $ \lesssim 10^{35}$ \\
  		Mimimal SUSY  (MSSM) SO(10)& $ \sim10^{34}$ \\
  		SUSY SO(10) & $ \sim   10^{32}-10^{35}$\\
  		Supergravity (SUGRA) SU(5) &$ \sim   10^{32}-10^{34}$ \\
  		Superstring (Flipped SU(5)) & $ \sim 10^{35}-10^{36} $\\
  		\hline
  	\end{tabular}
  	\caption{ Proton  half life  $(\tau_p)$ in various models \cite{adams2001proton,langacker1981grand,nath2007proton}.}
  \end{table}
  The GUT theories conserve the quantum number $ B-L$ instead of baryonic  $B$ and leptonic $L$ numbers individually. Allowing a decay channel for the proton into a Lipton and anti quark, for example, the decay into a neutral pion and positron via the channel
 \begin{equation*}
 p \longrightarrow\pi^0 + e^+
 \end{equation*}
 Which is the most famous decay channel. Nevertheless, the decay channel $ p \rightarrow\pi^0 + \mu^+$ is also predicted, particularly from monopoles \cite{sreekantan1984searches}. The table \ref{protonHL} summarizes the proton half life for different theories. We observe that there is a Large gap between the half-lives of proton predicted by $D=4$ quantum gravity and GUT and SUSY models. The existence of higher dimensions could bring the half lives closer, due to the effect of higher dimensional quantum gravity models on the quantum gravity mass $ M_{qg}$. Considering the GUP deformation as a phenomenological model of quantum gravity, it would be interesting to investigate the GUP deformation on the quantum gravity mass $ M_{qg}$ and hence the proton  half life. 

\section{Virtual black holes and uncertainty principle}
It is known from the uncertainly principle \`{a} la Heisenberg that in order to localise a system with accuracy $ \Delta x$ one needs to shine it with a photon with wavelength $ \lambda$ 
\begin{equation}
\Delta x \sim \lambda.
\end{equation}
{However, since that photon carries momentum inversely proportional to its wavelength i.e. $ p = \frac{h}{\lambda}$, we obtain the well-known uncertainty principle,
\begin{align}
\Delta x \, \lambda ^{-1} \sim 1 \\ \nonumber
\Delta x \Delta p \sim h \sim \hbar.
\end{align}
From the standard dispersion relation of the photon momentum and energy $ E= p $, it is obvious that localisation of a system to short distances adds significant amounts of energy to that system. If that system is the spacetime, we expect a gravitational back reaction to the energy added to the spacetime from Einstein field equations:
 \begin{equation}
 G_{\mu \nu} = 2\kappa T_{\mu \nu}
 \label{efe}
 \end{equation}
 We may approximate the~ \eqref{efe} over a small region of spacetime of radius $L$ to obtain~\cite{adler1999gravity}
 \begin{equation}
\frac{\delta g_{\mu \nu}}{L^2} \sim \frac{8 \pi G}{L^3} E.
 \end{equation}
 Where $\delta g_{\mu \nu}$ is the deviation from the flat metric, and it is given by the fractional uncertainty in positions $ \Delta x/L$ . We also identify Schwarzschild radius as $2MG$, with $ E = M$. We obtain the uncertainty relation :}
\begin{equation}
\Delta x \delta r_s \sim \ell_p^2
\end{equation}
Leasing to the conclusion that one cannot localise a region of the spacetime more to a radius of Plank length without causing a virtual black hole to form~ \cite{faizal2012some,hawking1996virtual} forming a ` quantum' foam structure for the spacetime at the Plank scale~\cite{scardigli1997black}. {This is a direct result from the HUP and not connected to any particular model of quantum gravity, including the GUP.}

\section{Virtual black holes and GUP}
{Since the formation of virtual black holes is a direct result for the marriage between the  Heisenberg uncertainty principle and general relativity.  Deforming the Heisenberg uncertainty principle (HUP) , to a generalised uncertainty principle (GUP), will affect the nature of the virtual black holes and their physical properties, for further reading about the relation between virtual black holes and GUP c.f. \cite{Scardigli:1999jh,adler2001generalized}.
We are mainly concerned of calculating the minimal mass for black holes using GUP, this will correspond to the mass of virtual black holes appearing from GUP instead of HUP. Since the mass of the virtual black holes is  what is used in calculating the proton half life.}
In order to compute the minimal black hole mass in GUP, we follow a similar argument made in \cite{ali2012no,cavaglia2004classical,medved2004conceptual}.  We make the argument very general and consider $D$ dimensional spacetime with all the$D-1$ momenta $p_i$ being equal  and  the quadratic GUP is given by \cite{cavaglia2003will,medved2004conceptual}
\begin{equation}
\Delta x \Delta p \geq \frac{\hbar}{2}\left( 1+14.9\left( \frac{D-3}{\pi}\right) ^2 \alpha^2 \frac{\Delta p^2}{M_p^2}\right) 
\label{gupq}
\end{equation}
Which leads to expression for $ \Delta p$
\begin{equation}
\Delta p \geq \frac{\Delta x}{\xi \, \alpha^2}\left\lbrace 1-\sqrt{1-\frac{\xi \hbar \alpha^2}{\Delta x^2}} \right\rbrace ,
\end{equation}
where $ \xi = 14.9\left( \frac{D-3}{\pi}\right) ^2$.
Now, let a particle be bounded at the black hole event horizon $ \Delta x \sim r_s$, this particle resembles a particle emitted by Hawking radiation from the horizon at Temperature associated with the resulting uncertainty in the momentum/energy of that particle localized at the horizon. Therefore the modified Hawking temperature is calculated using this argument in~\cite{ali2012no,scardigli2017gup,alsaleh}
\begin{equation}
T_{QGUP} = 2 T_H \left[ 1+ \sqrt{1-\frac{\xi \alpha_0^2}{4\lambda_D^2 m ^{\frac{2}{D-3}}}}\right] 
\end{equation}
In which
\begin{gather}
\lambda_D =\left( \frac{16 \pi}{(D-2)\Omega_{D-2}}\right) ^{\frac{1}{D-3}} \nonumber \\
\Omega_D = \frac{2 \pi^{\frac{D-1}{2}}}{\Gamma(\frac{D-1}{2})} \\
m= \frac{M}{M_p \nonumber}
\end{gather}
The GUP modified temperature is only physical for particular masses i.e.
\begin{equation}
\xi \alpha_0^2 \leq 4\lambda_D^2 m ^{\frac{2}{D-3}}
\end{equation}
The minimal mass $M_{min}$ of which the inequality above becomes equality is the mass of GUP virtual black holes, and it is given by
\begin{equation}
M_{QGUP} ( \alpha) =  M_p\left( \sqrt{\frac{ \pi\xi}{4}}\right) ^{D-3} \;  \frac{D-2}{8\Gamma(\frac{D-1}{2})}  \; \alpha ^{D-3}
\end{equation}
Since what is multiplied with $ \alpha$ is a pure numerical factor that depends on the dimension of spacetime, we denote such factor by $ f(D)$ as a function of spacetime dimensions, and the $D$ dimensional Plank mass $M_p^{(D)}$,
\begin{equation}
M^{(D)}_{QGUP} = f(D) M^{(D)}_p  \alpha ^{D-3}
\end{equation}
The proton half life is given by the expression from which we can calculate the minimal mass of the black holes in QGUP~$ M_{QGUP}$ using $ f(D)$, see table~\ref{f} :
\begin{equation}
\tau_p \sim M_{proton}^{-1} \left( \frac{M^{(D)}_p}{M_{proton}}\right) ^D \; f(D)^D  \alpha ^{D(D-3)}
\label{f}
\end{equation}
\begin{table}[h!]
\centering
\begin{tabular}{|c|c|c|}
\hline
$D$ & $f(D)$ & $ f(D)^D$ \\
\hline
$4$ & $0.307173$& $0.00890293$\\
$6$ & $13.1118$& $5.08128\times 10^6$ \\
$9$ & $11342.1$& $3.10629\times 10^{36}$ \\
$10$ & $128517$& $1.22912\times 10^{51}$\\
\hline
\end{tabular}
\caption{Different values of the parameter $f(D)$ and $ f(D)^D$ that appears in the proton half life formula, for different spacetime dimensions.  }
\label{f}
\end{table}
\section{Linear and quadratic GUP }
Another possible, and more general generalized uncertainty relation is \cite{ali2009discreteness,ali2012no}
\begin{equation}
\Delta x\Delta p \geq \frac{\hbar}{2}\left\lbrace  1- \alpha\, \frac{3.76}{M_p}  \left( \frac{D-3}{\pi }\right) \Delta p+\alpha^2 \, \,\frac{ 7.64}{M_p^2} \left( \frac{D-3}{\pi }\right)^2 \Delta p^2\right\rbrace 
\label{gupl}
\end{equation}
This uncertainty relation can be used similar to \eqref{gupq} to find the mass of GUP-deformed virtual black holes and study proton decay using the same argument as the previous section. \\
Solving \eqref{gupl} for $ \Delta p$.
\begin{equation}
\Delta p \geq \frac{2\Delta x + \alpha M_p \gamma_1}{\alpha^2 \gamma_2}\left( 1-\sqrt{1-\frac{2 \gamma_2 \alpha^2 M_p^2}{2\Delta x + \alpha M_p \gamma_1}}\right) 
\end{equation}
Such that
\begin{gather}
\gamma_1 = 3.76  \left( \frac{D-3}{\pi }\right) \nonumber \\
\gamma_2 = 15.28 \left( \frac{D-3}{\pi }\right)^2
\end{gather}
The modified Hawking temperature is then found to be 
\begin{equation}
T_{LQGUP} = 2T_H\left( 1+\frac{\gamma_1}{4 \lambda_D m^{\frac{1}{D-3}}}\right) ^{-1}\left\lbrace 1+\sqrt{1-\frac{\gamma_2 \alpha^2}{8(\lambda_D m^{\frac{1}{D-3}}+\frac{1}{4} \alpha \gamma_1)^2}}\right\rbrace ^{-1}
\end{equation}
The GUP modified temperature is only physical for particular masses i.e.
\begin{equation}
\gamma_2 \alpha^2 \leq 8(\lambda_D m^{\frac{1}{D-3}}+\frac{1}{4} \alpha \gamma_1)^2
\end{equation}
Leading to the minimal mass, $M_{LQGUP}$ that is the mass of virtual black holes
\begin{equation}
M_{LQGUP}(\alpha) = M_p\left(\frac{1}{2}\sqrt{ \frac{\gamma_2}{2}}-\frac{\gamma_1}{4}\right)^{D-3} \; \frac{D-2}{8 \Gamma(\frac{D-1}{2})} \pi^{\frac{D-3}{2}}\, \alpha^{D-3}.
\end{equation}
Since what is multiplied with $ \alpha$ is a pure numerical factor that depends on the dimension of spacetime, just like the quadratic GUP, we denote that factor by $ g(D)$, whose values are give in table~\ref{g} 
\begin{equation}
M^{(D)}_{GUP} = g(D) M^{(D)}_p  \alpha ^{D-3}
\end{equation}
The proton half life is given by the expression :
\begin{equation}
\tau_p \sim M_{proton}^{-1} \left( \frac{M^{(D)}_p}{M_{proton}}\right) ^D \; g(D)^D  \alpha ^{D(D-3)}
\label{g}
\end{equation}
\begin{table}[h!]
	\centering
	\begin{tabular}{|c|c|c|}
		\hline
		$D$ & $g(D)$ & $ g(D)^D$ \\
		\hline
		$4$ & $0.369562$& $0.0186531$\\
		$6$ & $22.8336$& $1.41726\times 10^8$ \\
		$9$ & $34396.9$& $6.74015\times 10^{40}$ \\
		$10$ & $468909$& $5.13913\times 10^{56}$\\
		\hline
	\end{tabular}
	\caption{Different values of the parameter $g(D)$ and $ g(D)^D$ that appears in the proton half life formula, for different spacetime dimensions.  }
	\label{g}
\end{table}
We can use the data for Planck masses in different spacetime dimensions $M_p^{(D)}$ \cite{chatrchyan2011search}  and the relations \eqref{f}\eqref{g} to estimate the bounds on the GUP deformation parameter    using quadratic in $p$  $ \alpha$ and linear-quadratic in $p$ deformations $ \alpha' $.
Knowing that the experimental bound on the proton half-life is $ >10^{34}$ years \cite{sreekantan1984searches,PhysRevLett.102.141801} and the proton mass $ \sim 10^3$ GeV.
The bound on $ \alpha$ is given by:
{
\begin{align}
10^{34} &> 10^{-31} \,10^{-3D} M_p^{(D)D} g(D)^D \alpha^{D(D-1)} \nonumber \\
10^{65-3D} &> 1 M_p^{(D)D} g(D)^D \alpha^{D(D-1)}\nonumber \\
10^{\frac{65+3D}{D(D-1)}} M_p^{(D)-\frac{1}{D-1}} g(D)^{-\frac{1}{D-1}} &>   \alpha
\end{align}
}
Or
{
\begin{equation}
 \alpha_{LQGUP} < 10^{-\frac{65+3D}{D(D-1)}} M_p^{(D)-\frac{1}{D-1}} g(D)^{-\frac{1}{D-1}}.
\end{equation}
Same goes for quadratic GUP,
\begin{equation}
 \alpha_{QGUP} < 10^{\frac{65+3D}{D(D-1)}} M_p^{(D)-\frac{1}{D-1}} f(D)^{-\frac{1}{D-1}}.
\end{equation}
}
Table~\ref{mass} contains the values for minimal black hole masses for both QGUP and LQGUP in different spacetime dimensions. Moreover, table \ref{alpha} contains the bounds on $  \alpha_{QGUP}$ and $  \alpha_{LQGUP}$  in quadratic and linear-quadratic  GUP deformations in different spacetime dimensions, relevant to physical models. It appears that the GUP deformations become more appearent in the low energy limit if the spacetime has extra dimentions. 
\begin{table}[h!]
	\centering
	\begin{tabular}{|c|c|c|c|}
		\hline
		$D$ & $M_p^{(D)}$ GeV & Quadratic GUP $M_{QGUP}$ GeV &Linear-Quadratic GUP $M_{LQGUP}$ GeV\\
		\hline
		$4$ & $1.22 \times 10^{19}$& $3.66 \times 10^{18}$&$4.4 \times 10^{18}$\\
		$6$ & $4.54 \times 10^{3}$& $5.95\times 10^4$& $1\times 10^5$ \\
		$9$ & $2.71 \times 10^3$& $3.1\times 10^{7}$&$9.5\times 10^{7}$ \\
		$10$ & $2.51 \times 10^3$& $3.22\times 10^{8}$&$1.17\times 10^{9}$\\
				\hline
	\end{tabular}
	\caption{Planck masses in different spacetime dimensions, compared with the minimal black hole masses in quadratic and linear-quadratic GUP.}
	\label{mass}
\end{table}
\begin{table}[h!]
	\centering
	\begin{tabular}{|c|c|c|}
		\hline
		$D$ & Bound on quadratic $\alpha_{}$ & Bound on Linear-quadratic $\alpha$ 
	\\
		\hline
		$4$ & $<4.37\times 10^{-3}$& $<3.64\times 10^{-3}$\\
		$6$ & $<101.01$& $< 84.95$ \\
		$9$ & $<0.87$& $<0.74$ \\
		$10$ & $<0.51$& $<0.42$\\
		\hline
	\end{tabular}
	\caption{{Upper} bounds on the GUP deformation parameters $  \alpha_{QGUP}$ and $ \alpha_{LQGUP}$. From the experimentally measured half-life of the proton. }
	\label{alpha}
\end{table}

\section{Conclusions}

In this work, we investigated the production of virtual black holes in higher dimensions in the context of generalized uncertainty principle. We use this black hole production to study the proton decay process that is considered  as a mediator for these virtual black holes. We calculate the proton half life in higher dimensions and we set an {upper} bound on the GUP deformation parameter $ \alpha_{QGUP}$ and $ \alpha_{LQGUP}$. We found that the bounds on GUP parameters are around 100, .87, and 0.51 for 6, 9 and 10 dimensions, respectively. These values are stringent and consistent with the bound set by electroweak scale \cite{Ali:2011fa}. In fact, this is  an improvement
for various studies on phenomenological aspects of GUP, 
if the GUP parameter $\alpha \sim 1$, it appears to be a new and interesting result and relevant to be studied at low energy systems\cite{Ali:2011fa,Das:2008kaa,Scardigli:2014qka} and as appeared from comparing the thermal corrections to black hole temperature with temperature from quantum-corrected Newtonian potential~\cite{scardigli2017gup,alsaleh} . This indicate that GUP could be useful to explain the proton decay process beyond the standard model and could open an interesting phenomenological window for studying quantum gravity effects for low energy systems. We hope to report on these issues in the future. 
\section*{Acknowledgements}
We would like to thank the anonymous referees for their useful comments and help improving this manuscript.\\
This research project was supported by a grant from the " Research Center of the Female Scientiffic and Medical Colleges ", Deanship of Scientiffic Research, King Saud University. 
\bibliographystyle{plain}
\bibliography{ref.bib}
\end{document}